\documentclass[12pt,preprint]{aastex}
\usepackage{epstopdf}

\begin{document}

\title{Imaging the Molecular Disk Orbiting the Twin Young Suns of V4046 Sgr} 

\author{David R. Rodriguez\altaffilmark{1}, Joel H.\
  Kastner\altaffilmark{2}, David Wilner\altaffilmark{3}, \& Chunhua
  Qi\altaffilmark{3} }

\altaffiltext{1}{Dept.\ of Physics \& Astronomy, University of California, Los Angeles 90095, USA \\
(drodrigu@astro.ucla.edu)}
\altaffiltext{2}{Center for Imaging Science, Rochester Institute of
Technology, 54 Lomb Memorial Drive, Rochester NY 14623
(jhk@cis.rit.edu)}
\altaffiltext{3}{Harvard-Smithsonian Center for Astrophysics, 60 Garden Street, Mail Stop 42, Cambridge, MA 02138 (dwilner@cfa.harvard.edu \& cqi@cfa.harvard.edu)}

\begin{abstract}
  We have imaged the disk surrounding the nearby ($D\sim73$~pc),
  $\sim$12~Myr, classical T Tauri binary system V4046 Sgr with the
  Submillimeter Array (SMA) at an angular resolution of
  $\sim$2\arcsec. We detect a rotating disk in $^{12}$CO(2--1) and
  $^{13}$CO(2--1) emission, and resolve the continuum emission at 1.3
  mm. We infer disk gas and dust masses of $\sim$110 and $\sim$40
  Earth masses, respectively. Fits to a power-law disk model indicate
  that the molecular disk extends to $\sim$370 AU and is viewed at an
  inclination of between $\sim$33$^\circ$ and $\sim$39$^\circ$
    for dynamical stellar masses ranging from 1.8 $M_\odot$ down to
    1.5 $M_\odot$ (the range of total mass previously determined for
    the central, $2.4$~day spectroscopic binary). This range of disk
    inclination is consistent with that assumed in deducing the
    central binary mass (i.e., 35$^\circ$), suggesting that the V4046
    Sgr binary system and its circumbinary, molecular disk are
    coplanar.  In light of the system's age and binarity, the
  presence of an extensive molecular disk orbiting V4046 Sgr provides
  constraints on the timescales of processes related to Jovian planet
  formation, and demonstrates that circumbinary Jovian planets
  potentially could form around close binary systems.
\end{abstract} 

\keywords{circumstellar matter --- stars: binaries, individual (V4046 Sgr) --- stars: pre-main sequence --- stars: planetary systems --- radio lines : stars}

\section{Introduction}

Circumstellar disks around young stars serve both as the sources of
material for accreting young stars and as the sites of nascent planets
orbiting such stars. Numerous studies of such disks have exploited photometry
and spectroscopy of dust emission, as manifest in the form of a
thermal infrared excess above the stellar continuum, to ascertain
fundamental properties such as disk dimensions, density structure,
dust mass, and dust grain composition \citep[e.g.,][and references
therein]{z01,dullemond07}. Observations that can establish the
composition and evolution of the {\it gaseous} component within
circumstellar disks around pre-MS and young MS stars,
meanwhile, are essential if we are to understand the processes
involved in Jovian planet formation and the origins of comets and
Kuiper Belt objects \citep{lj02,hubbard02,ec00}. For
example, the near-IR lines of rovibrational H$_2$ and CO generally
probe the terrestrial planet formation zones ($\stackrel{<}{\sim}$2
AU) of gas disks, interior to the regions where Jovian planets
are expected to form 
(e.g., \citealt{Salyk07,Najita08,Bary08}; however, see \citealt{Brittain09}). Mid-IR pure rotational
lines of H$_2$ \citep{Bitner07,Bitner08,Najita10} and
rovibrational lines of H$_2$O, C$_2$N$_2$, and OH 
\citep[and references therein]{Glassgold09} are potentially diagnostic of molecular
gas mass and excitation in the innermost Jovian planet-forming zones,
and also provide limited information concerning the structure,
dynamics, and chemistry of these regions.

Sensitive measurements of (sub)millimeter-wave radio emission from orbiting CO (after H$_2$, the most abundant molecular species) around ``isolated,'' young (age $\sim$5--$20$ Myr), nearby ($D\lesssim$ 100~pc) stars --- with TW Hya as archetype --- are key in this regard \citep{z95,k97,thi04}.
Such stars are old enough that Jovian planets may already be forming or have formed in their circumstellar disks \citep[e.g.,][]{pn05,thommes08}, and they are close enough for intensive follow-up study via interferometric molecular line imaging surveys.
Though CO detections of the disks around pre-main sequence stars of such advanced ages are exceedingly rare thus far, (sub)millimeter-wave molecular imaging studies of the handful of known $\sim$10~Myr-old molecular disks that lie within $\sim$100~pc of Earth have yielded a wealth of information concerning the structure and evolution of gaseous, planet-forming disks \citep[e.g.,][]{qi04,qi06,qi08,hughes08}.

V4046 Sgr is one such isolated, ``old,'' nearby star system \citep[age
$\sim$12~Myr, distance $\sim$73~pc;][]{torres08}.  The V4046 Sgr
system is especially noteworthy in that it consists of a close
($\sim$9~$R_\odot$ separation, 2.4-day period) pair of $\sim$0.9
$M_\odot$ stars that are evidently still actively accreting mass
\citep[][and references therein]{sg04}.  A previous molecular emission
line survey with the Institut de Radio Astronomie Millimetrique (IRAM)
30-m telescope established the presence of CO, HCN, CN, and HCO$^+$
within the dusty disk encircling the system \citep{k08b}.  This makes
V4046 Sgr only the fourth known example \citep[after TW Hya, 49 Cet,
and HD~141569;][]{z95} of a pre-main sequence star within $\sim$100~pc
that possesses a molecular disk; V4046 Sgr is the only known close-separation binary
among them (HD~141569 has a pair of M-type companions, 
$\sim$800~AU from the disk-bearing primary, 
see \citealt{weinberger00}). 
To confirm that the CO emission detected toward V4046 Sgr
with the IRAM 30-m indeed arises in an orbiting, circumbinary disk,
and to establish or constrain fundamental system parameters such as
disk dimensions and inclination, we observed V4046 Sgr with the
Submillimeter Array \citep[SMA;][]{ho04}.

\section{Observations}\label{observations}

We obtained data in the SMA's extended (28--226 m baselines) and
compact (6--70 m baselines) configurations on 23 Feb.\ and 25 April 2009, respectively. The weather conditions were excellent on both nights.
During each track, the receivers were tuned to provide simultaneous
coverage of the $^{12}$CO(2--1), $^{13}$CO(2--1), and C$^{18}$O(2--1)
transitions as well as adjacent (1.3 mm) continuum, with the spectral backends
configured to provide velocity resolution of 0.26~km~s$^{-1}$.  The
data were edited and calibrated using the MIR software package\footnote{See http://cfa-www.harvard.edu/$\sim$cqi/mircook.html}.
The passband response was calibrated using the quasars 3C~273 and
1924-292 for the extended configuration and 3C~84, 1924-292, and
Uranus for the compact configuration. The antenna gain calibration was
performed using 1924-292 (extended) and 1733-130 (compact).  The flux
scale was determined by bootstrapping observations of 3C~273 and
Uranus and is accurate at the $\sim$10\% level.
The continuum was determined using the line-free 
channels in both upper and lower sidebands for both configurations.
From these data we determine a 1.3~mm continuum flux of $360\pm36$ mJy.

We combined the extended and compact configuration data to produce a synthesized beam
size of $2.2''\times1.5''$ at the 230 GHz frequency of $^{12}$CO(2--1). Similarly, for $^{13}$CO(2--1) we produced a synthesized beam size of $2.3\arcsec \times 1.7\arcsec$ at its 220 GHz frequency.
No C$^{18}$O 219 GHz emission was detected with a 3$\sigma$ upper limit of 0.3~Jy beam$^{-1}$ (synthesized beam size was $2.4\arcsec \times 1.7\arcsec$). The standard tasks (CLEAN algorithm) of deconvolution and image restoration were performed with the MIRIAD software package.
Images of the continuum and $^{12}$CO were created using Briggs weighting (robust=0 and 1, respectively) for the visibilities.
To reduce image noise, the $^{13}$CO and C$^{18}$O
images were generated using natural weighting. The complete
observing parameters are listed in Table~\ref{tab1}.

\section{Results}\label{results}

\subsection{CO line maps}

\begin{figure}[htb]
\begin{center}
\includegraphics[width=18cm,angle=0]{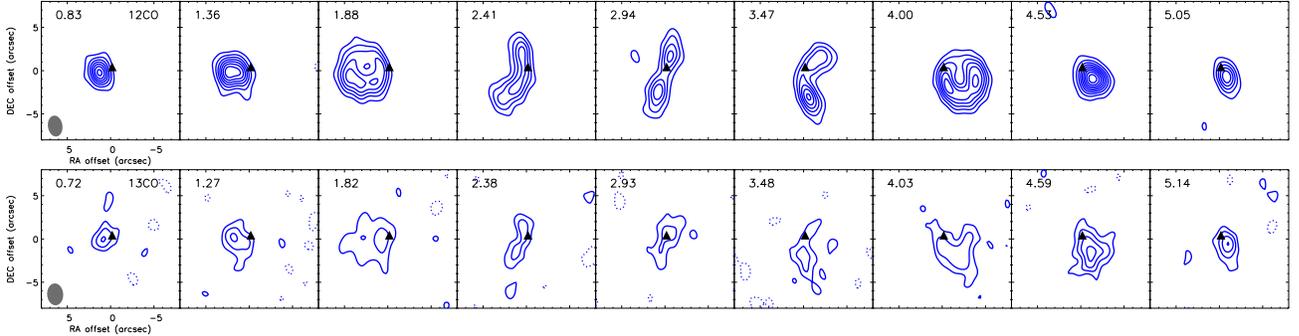}
\end{center}
\caption{SMA velocity-resolved maps of V4046 Sgr in $^{12}$CO(2--1) (top) 
and $^{13}$CO(2--1) (bottom).
The contour levels are 0.09~mJy/beam$\times$[3,6,9,...] for $^{12}$CO(2--1) and 
0.08~mJy/beam$\times$[$-2$,2,4,6,...] for $^{13}$CO(2--1) with dotted lines 
representing the negative contours.
  In each frame, the central velocity (in km s$^{-1}$ with respect t
  o the Local Standard of Rest (LSR)) is indicated at
  upper left; the position of the continuum source centroid is
  indicated by a small black triangle; and the synthesized beam size, shape, and
  position angle is indicated by the grey shaded ellipse at lower
  left.}
\label{fig:SMAmaps}
\end{figure}

The resulting sequences of velocity-resolved SMA $^{12}$CO(2--1) and
$^{13}$CO(2--1) images are displayed in Fig.~\ref{fig:SMAmaps}.  The
contours start at 3 and 2 times the rms noise level, respectively, and
increase as described in the figure caption. 
These images vividly demonstrate that the CO 
emission from V4046 Sgr arises from a rotating disk.
The radial extent of the
molecular disk is $\sim$5\arcsec, or $\sim$370~AU at the estimated
73~pc distance to V4046 Sgr.  The similar overall extent of the CO
emission in the N-S and E-W directions, coupled with the profound
change in emission morphology with radial velocity, indicates that the
disk is viewed at intermediate inclination (compare with, e.g., Fig.\
3 in \citealt{bs93}, which illustrates a Keplerian molecular disk
model viewed at an inclination of $\sim$45$^\circ$).  The channel map
closest to the systemic LSR velocity ($V_{sys} \sim 2.9$ km s$^{-1}$)
indicates a disk position angle (PA) on the sky of $\sim$75$^\circ$,
measured east of north. This PA is similar to that inferred for the
dust disk from the 230 GHz continuum emission detected and resolved by
the SMA (see \S~\ref{continuum}). We further refine these estimates
for disk inclination, PA, and $V_{sys}$ in \S~4.1.

\subsection{CO line profiles}

\begin{figure}[htb]
\begin{center}
\includegraphics[width=10cm,angle=0]{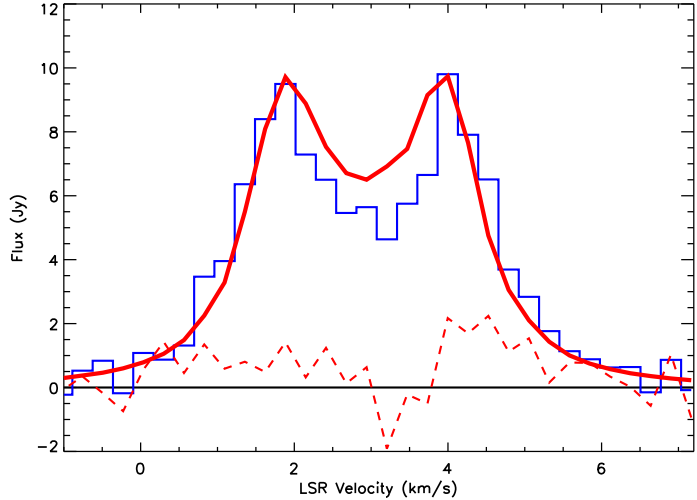}
\end{center}
\caption{Line profile of  $^{12}$CO(2--1) obtained from the SMA images of the V4046 Sgr disk. Overlaid in bold is the best-fit model as described in \S~\ref{models}. The dotted line shows the residual when the model is subtracted from the data in the visibilities.}
\label{fig:SMAspectra}
\end{figure}

\begin{figure}[htb]
\begin{center}
\includegraphics[width=10cm,angle=0]{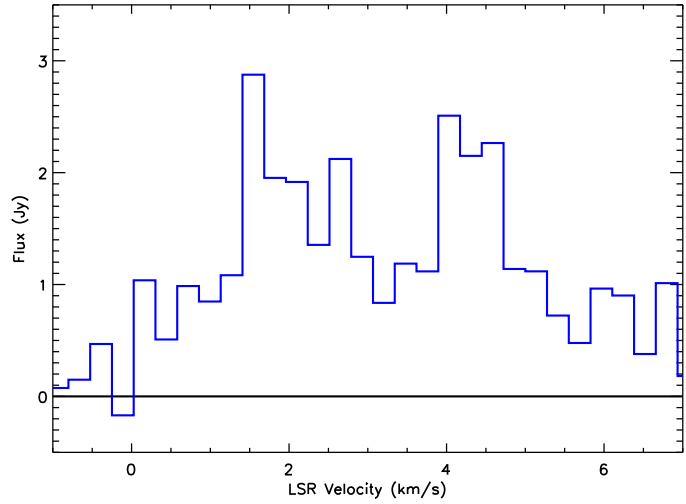}
\end{center}
\caption{Line profile of  $^{13}$CO(2--1) obtained from the SMA images of the V4046 Sgr disk.}
\label{fig:SMAspectra2}
\end{figure}

The SMA $^{12}$CO(2--1) and $^{13}$CO(2--1) emission line profiles, as
obtained by spatially integrating the line intensity maps within a
$6''$ radius circular aperture, are highly symmetric
(Figures~\ref{fig:SMAspectra} and~\ref{fig:SMAspectra2}). 
This confirms that the asymmetric
appearance of the CO line profiles obtained with the IRAM 30-m
telescope \citep{k08b} was a consequence of slight
mispointing (the 30-m data were obtained at very low source
elevation, $\sim$15--20$^\circ$, due to the southerly decination of
V4046 Sgr) combined with the spatial resolution of the source by the
30-m beam, as opposed to significant density or temperature
inhomogeneities intrinsic to the molecular disk.

\subsection{1.3 mm dust continuum emission\label{continuum}}

\begin{figure}[htb]
\begin{center}
\includegraphics[width=7cm,angle=0]{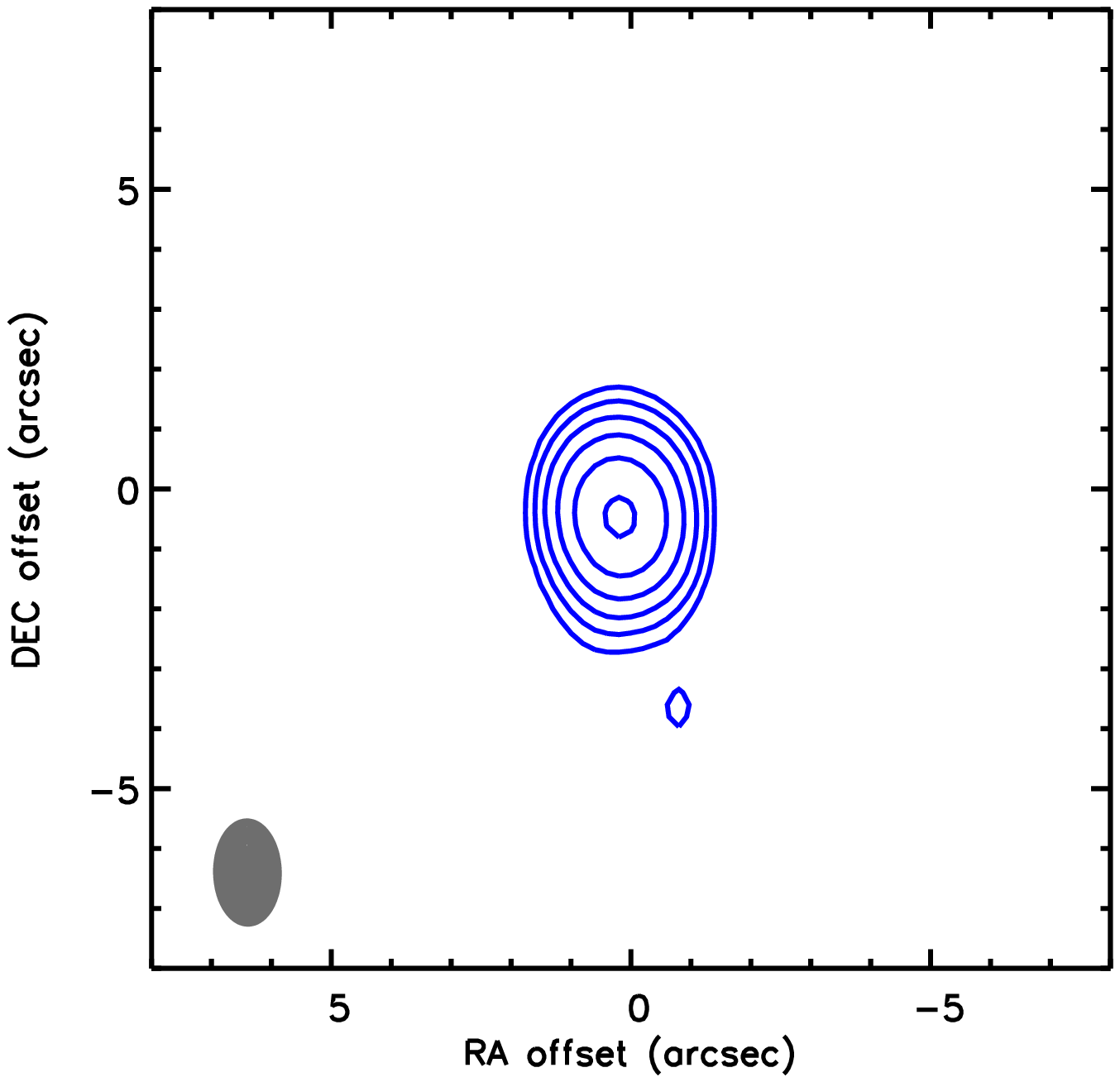}
\includegraphics[width=6.7cm,angle=90]{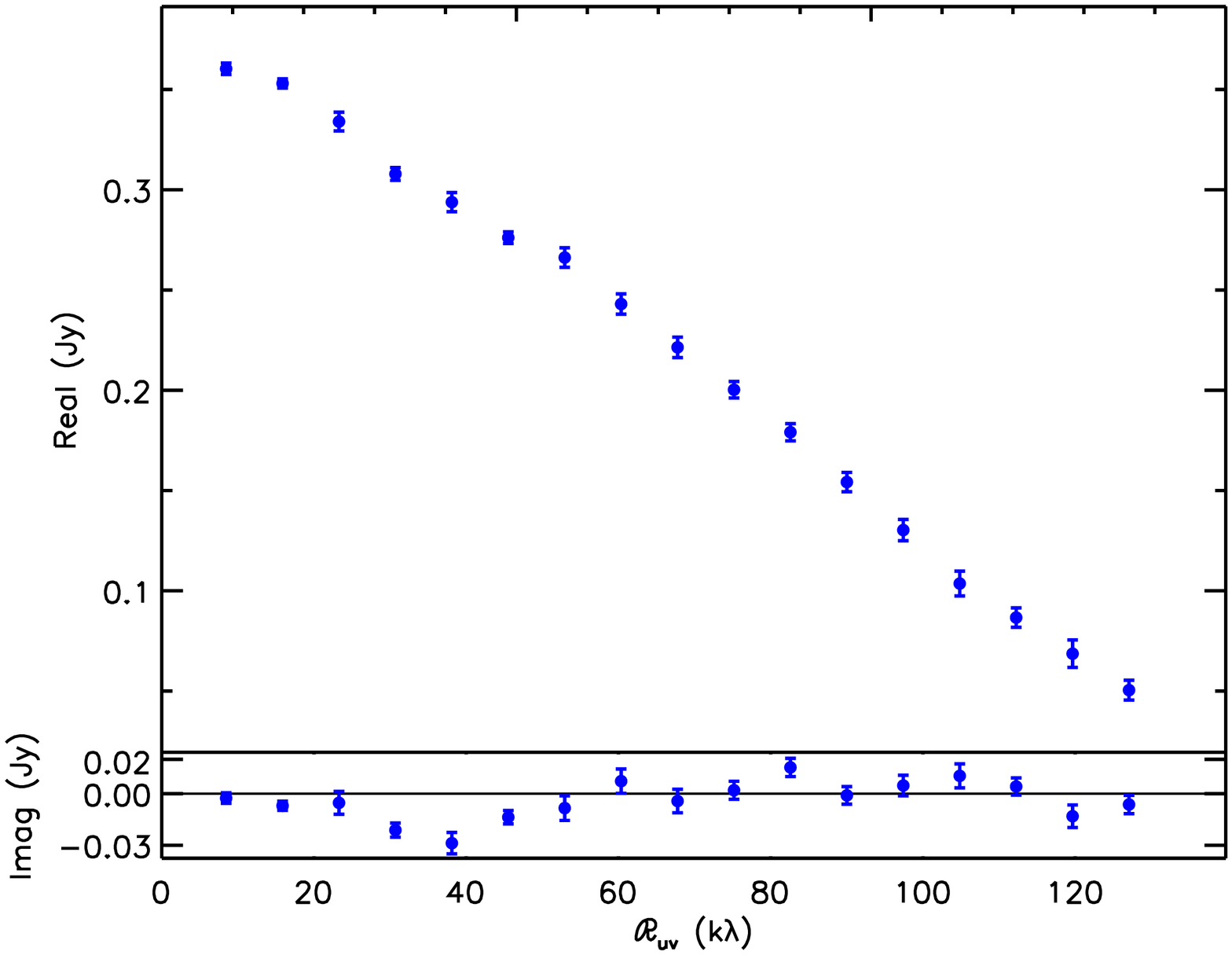}
\end{center}
\caption{Continuum emission map (left) and deprojected visibility plot (right). 
The decline in the real component of the visibility as a function of 
baseline is a characteristic of resolved structures. 
The FWHM corresponds to a radial extent of about 36~AU, consistent with the 
characteristic extent of 40~K dust grains in the system.}
\label{fig:cont}
\end{figure}

The continuum emission from the dust is partially resolved
(Figure~\ref{fig:cont}).  Using MIRIAD, we fit an elliptical gaussian
to the continuum visibilities to estimate the size and shape of the
emission. The best fit yields major and minor axes of $1.10\arcsec$
and $0.93\arcsec$, respectively, with a position angle of
$73\pm5^\circ$ east of north. This position angle is consistent with that
inferred from the CO maps (\S 4.1). There is negligible contribution
from the ($<0.1$\arcsec) seeing at 1.3 mm to this Gaussian fit. At a
distance of $\sim$73~pc, the disk resolved by the continuum would have
a characteristic radial extent of about 36~AU (corresponding to the
FWHM of the surface brightness distribution). 
Both the flux ($360\pm36$~mJy) and spatial extent of the disk at 1.3
mm are consistent with $\sim$40~K blackbody dust grains located at a
distance of $\sim$40~AU from the central binary 
(\citealt{k08b}\footnote{The value of $\sim$80~AU quoted in \citet{k08b} 
is actually a diameter. The semi-major axis is $\sim$40~AU.}).

\section{Modeling}\label{models}

\subsection{CO line maps}

To determine more precise values for the disk parameters, we employed
standard modeling techniques previously used to interpret
interferometric maps of molecular line emission
\citep[e.g.,][]{qi04,raman06,hughes08}. Specifically, we used the Monte Carlo
radiative transfer code RATRAN \citep{hv00} to solve for the CO level
populations at each position within the disk and thereby to generate
sky-projected emission-line images. These images were then passed to
the MIRIAD task {\it uvmodel} to sample the model images at the same
spatial frequencies as our SMA data. A minimum $\chi^2$ is obtained by
directly comparing the model and observed visibilities \citep{GD98}.
The model geometry is
that of a Keplerian disk in hydrostatic equilibrium with truncated
power laws in density and temperature described (respectively) by
$n(r) = n_{100} (\frac{r}{100AU})^{-p}$ and $T(r) = T_{100}
(\frac{r}{100AU})^{-q}$, where $n(r)$ refers to the number (mid plane)
density of H$_2$ and we assume CO:H$_2=10^{-4}$.

\begin{figure}[htb]
\begin{center}
\includegraphics[width=8cm,angle=0]{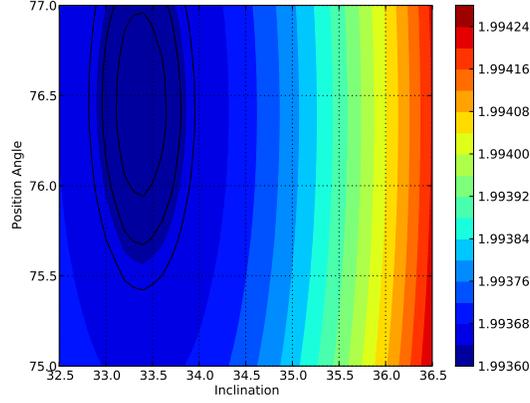}
\end{center}
\caption{Contour maps of reduced $\chi^2$ for a systemic velocity of 2.92 km s$^{-1}$ and outer radius of 370~AU. The black ellipses indicate 1-, 3-, and 5-$\sigma$ contours.}
\label{contourmaps}
\end{figure}

\begin{figure}[htb]
\begin{center}
\includegraphics[width=18cm,angle=0]{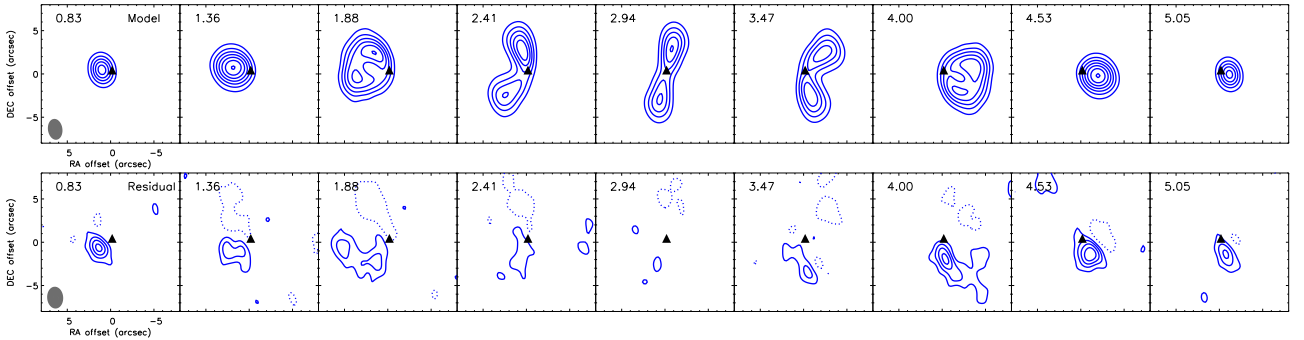}
\end{center}
\caption{$^{12}$CO(2--1) best-fit model map (top) and residual map (bottom).
  In each frame, the central velocity (km s$^{-1}$) is indicated at
  upper left; the position of the continuum source centroid is
  indicated by a small black triangle; and the synthesized beam size 
  ($2.3\arcsec\times1.6\arcsec$ at 8.5$^\circ$) is indicated by the grey shaded ellipse at 
  lower left.
  Contours levels are 0.09~mJy/beam$\times$[$-3$,3,6,9,...] for both panels 
  with dotted contours for the negative levels.
  Figure~\ref{fig:SMAspectra} illustrates how the residual flux varies over velocity.
  The integrated line flux of the residuals is $\sim$14\% of the $^{12}$CO and model fluxes.}
\label{fig:Modelmaps}
\end{figure}

\begin{figure}[htb]
\begin{center}
\includegraphics[width=8cm,angle=0]{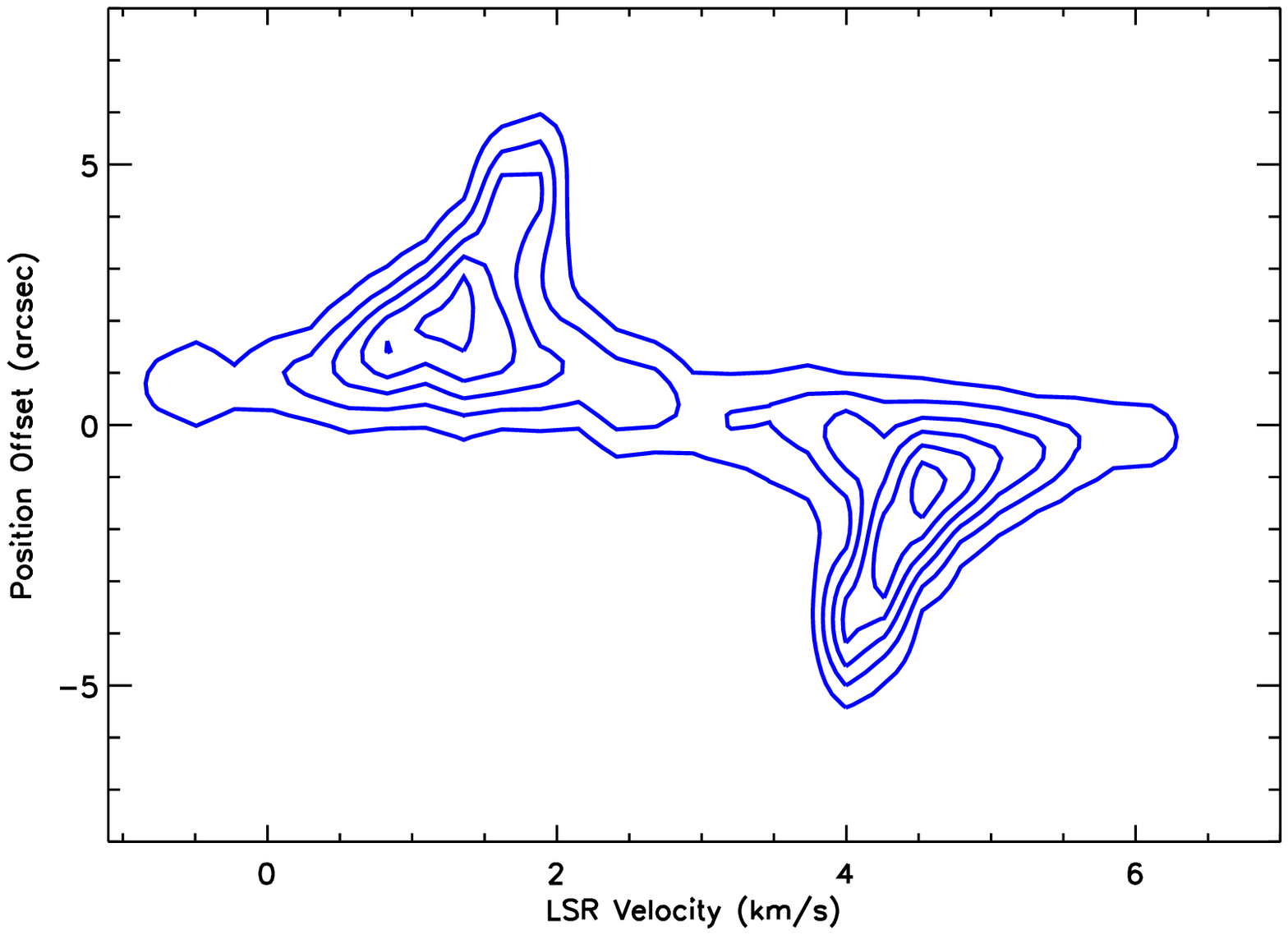}
\includegraphics[width=8cm,angle=0]{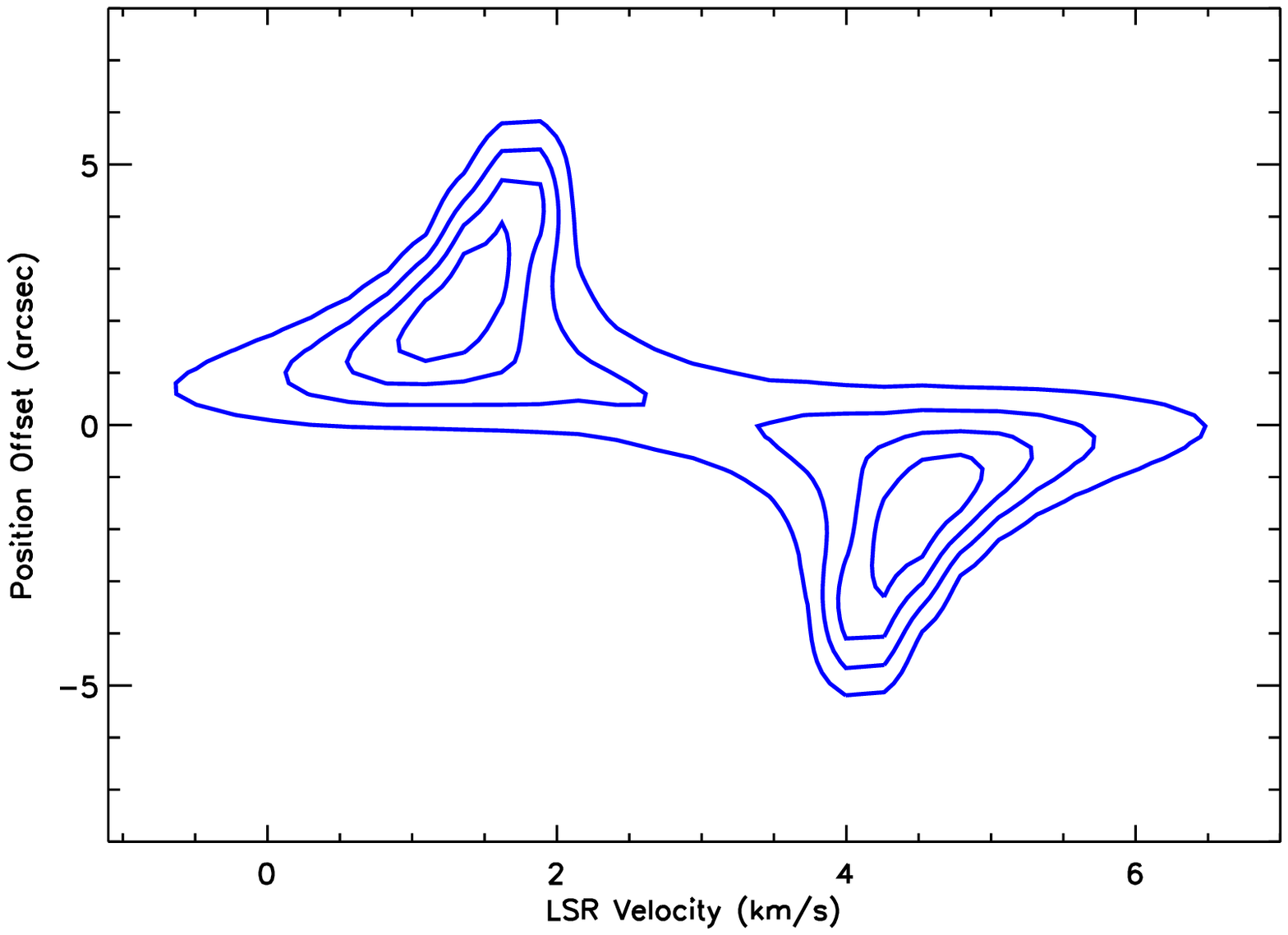}
\end{center}
\caption{Position-velocity diagram for the $^{12}$CO(2--1) emission in 
V4046 Sgr (left) and the best-fit Keplerian model (right).
Both diagrams were constructed for a P.A. cut of 76.4$^\circ$ through the 
position of the central binary star.
}
\label{fig:PV}
\end{figure}

There are ten parameters that describe this Keplerian disk model (see
Table 2). In
practice, however, our modeling can constrain only five of these: the
central mass, the systemic velocity ($V_{LSR}$), and the CO disk
inclination ($i$), PA, and outer truncation radius
($R_{out}$). The modeling is relatively insensitive to the other
parameters ($T_{100}$, $q$, $n_{100}$, $p$, and inner truncation
radius $R_{in}$) because the $^{12}$CO(2--1) emission imaged by the
SMA is optically thick \citep{k08b} and originates from the cooler,
outer regions of the disk (the $^{13}$CO(2--1) emission-line data are
too noisy to provide useful constraints on the disk density structure
parameters $n_{100}$ and $p$). Furthermore, there is 
degeneracy between central mass and disk inclination; for the former,
we initially adopted the total stellar mass of 1.8$M_\odot$ determined
by \citet{sg04}, then varied this parameter so as to explore the
potential range of $i$ in more detail (\S 4.2). The values of
parameters $T_{100}$, $q$, $n_{100}$, $p$, and $R_{in}$ (Table 2) were
fixed according to results obtained from models of the molecular disks
orbiting other T Tauri stars \citep{dutrey94,dutrey98}. The remaining
model parameters ($V_{LSR}$, P.A., $i$, $R_{out}$) were then varied in
turn so as to ascertain best-fit values. The resulting
set of best-fit model parameters and their formal 1-$\sigma$ uncertainties are
listed in Table~\ref{tab2}. These uncertainties do not account for
systematic errors (see below), as they were determined from the change
in reduced $\chi^2$ as the model parameters were varied; a
representative contour plot of reduced $\chi^2$ (illustrating its
dependence on P.A. and $i$) is presented in Figure~\ref{contourmaps}. 
The best-fit Keplerian disk model is presented in the form of
synthetic velocity-resolved images in Fig~\ref{fig:Modelmaps} and as a
position-velocity (P-V) diagram alongside the observed P-V diagram in
Fig~\ref{fig:PV}. 

The value of $R_{out} = 370\pm30$ AU obtained from the model fitting
confirms our earlier ``eyeball'' estimate. The systemic velocity we
find from fitting the SMA data, $2.92\pm0.01$~km~s$^{-1}$, is
consistent with that determined from single-dish, mm-wave CO
spectroscopy by \citet{k08b}. The inferred disk inclination of
$33.4\pm0.3$$^\circ$ is very similar to the value previously assumed
in determining the masses of the components of the central binary
system (see below). The position angle obtained from the best-fit
model ($76.4\pm0.5^\circ$) is consistent with that derived from the
elliptical gaussian fit to the continuum visibility
($73\pm5^\circ$). The truncated power law model generated by these
best-fit parameters reproduces well the overall emission morphlogies
apparent in the velocity channel maps. There is some structure in the
fit residuals (Fig~\ref{fig:Modelmaps}), however. In particular, the
data show an excess of emission from the southern portion of the disk
relative to the model. The integrated line flux of the residuals is
$\sim$14\% of the $^{12}$CO and model fluxes (see
Fig.~\ref{fig:SMAspectra}). Given these systematics, and those due to
the simplifying assumptions inherent in the molecular disk model, the
formal errors listed in Table~\ref{tab2} likely somewhat underestimate
the uncertainties in the best-fit model parameters.

\subsection{Central binary system mass vs.\ disk inclination}\label{massvary}

Estimates of the total mass of the V4046 Sgr close binary, based on
radial velocity curves obtained from optical spectroscopy of the
double-lined system, range from 1.55~M$_\odot$ \citep[component masses
0.86~M$_\odot$ and 0.69~M$_\odot$;][]{quast00} to 1.78 M$_\odot$
\citep[component masses 0.91~M$_\odot$ and 0.87~M$_\odot$;][]{sg04}.
Both estimates are based on an assumed binary system inclination of
35$^\circ$ \citep{quast00}. The \citet{sg04} component masses appear
to be more accurate, given the nearly equal radial velocity amplitudes
of the two components and the improved agreement with pre-main
sequence evolutionary tracks \citep[see disussion in][]{sg04}.

\begin{figure}[htb]
\begin{center}
\includegraphics[width=8cm,angle=0]{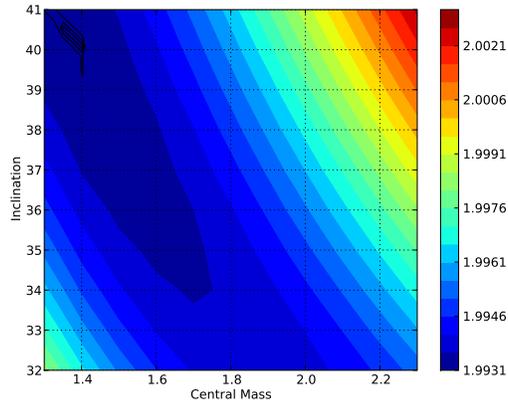}
\end{center}
\caption{Contour plot of reduced $\chi^2$ for disk inclination vs.\
  central (stellar) mass. 
  The black contours indicate 1-, 3-, and 5-$\sigma$ contours; 
  the jagged appearance is due to the coarseness of the grid used.
  The range of central masses covered in the
  plot is significantly larger than that determined via studies of
  stellar radial velocities (1.55--1.8 $M_\odot$; \citealt{quast00,sg04}) 
  under the assumption of a binary
  inclination of $35^\circ$.}
\label{ms_incl}
\end{figure}

To ascertain the constraints placed by the SMA CO maps on both the
mass of the central binary and the inclination of the disk, we fixed
the remaining disk model parameters to their best-fit
(Table~\ref{tab2}) values, and allowed only central mass and
inclination to vary. The results for reduced $\chi^2$, displayed in
Fig.~\ref{ms_incl}, illustrate the inherent degeneracy between mass
and inclination; lower central masses and higher inclinations yield
fits that are comparable in quality to those for higher central masses
and lower inclinations. Nonetheless, this fitting exercise
demonstrates that, for a central mass in the range 1.55 to
1.8~M$_\odot$ allowed by the optical spectroscopy, the disk
inclination is constrained to values between 33$^\circ$ and
37$^\circ$. Given the similarity of these values of disk inclination
to that assumed for the central binary in determining component masses
from stellar radial velocities, it appears that the molecular disk is
nearly coplanar with the central binary system. Although smaller
central binary masses (and hence larger disk inclinations) are not
precluded by our disk model fitting --- indeed, we find that $\chi^2$
reaches a minimum near $i\sim40^\circ$, for a central mass of
$\sim$1.4~M$_\odot$ --- values of central mass $\le1.5$~M$_\odot$ for
this system lead to discord with pre-main sequence evolutionary tracks
\citep{sg04}.

\subsection{Disk gas and dust masses}

 To estimate the disk molecular mass 
  from the integrated SMA $^{12}$CO and
  $^{13}$CO line intensities (30.2 Jy km s$^{-1}$
  for $^{12}$CO and 8.7 Jy km s$^{-1}$ for $^{13}$CO) 
  we adopt the methods and assumptions described in \citet{k08a,k08b}
  and references therein. In particular,
  we assume optically thin $^{13}$CO emission and 
  CO:H$_2$ and $^{12}$C:$^{13}$C ratios of $10^{-4}$ and 89, respectively;
  the latter values are highly uncertain.
  We assume the mean gas temperature is similar to that of the dust (37~K; see below).
  The ratio of the $^{12}$CO and $^{13}$CO line intensities indicates a
  $^{12}$CO optical depth of $\sim$26.
  We use this result to infer a disk molecular mass of
  $\sim$110 Earth masses from the $^{12}$CO line intensity \citep[see Equation 4 in][]{z08}.
  This gas mass estimate is consistent with that obtained for V4046 Sgr on the
  basis of single-dish (IRAM 30-m) data 
  (\citealt{k08b}\footnote{Due to a numerical transcription error, 
  the gas mass estimate of $\sim$13~Earth masses stated in \citet{k08b} is an 
  order of magnitude too small.}).
  We note that the assumption that the
  $^{13}$CO emission is optically thin is supported by the
  nondetection of C$^{18}$O(2--1) emission by the SMA, although 
  the large $^{12}$CO optical depth places a lower limit
  on the $^{13}$CO optical depth of $\sim$0.4. 

  Adopting a dust temperature of 37 K --- obtained from a blackbody
  fit to the far-infrared/submillimeter
  spectral energy distribution of V4046 Sgr \citep{k08b}
  --- and a dust
  opacity of 1.15 cm$^2$ g$^{-1}$ (extrapolated from an opacity of 1.7
  cm$^2$ g$^{-1}$ at 880~$\mu$m assuming a dust opacity slope $\beta$=1; see Zuckerman et al.\ 2008), we
  estimate a disk dust mass of $\sim$40 Earth masses. This is a factor
  of $\sim$2 larger than the estimate obtained by \citet{k08b};
  the discrepancy is likely due to the adopted power-law dependence of opacity
  on wavelength.

\section{Summary and Conclusions}

Our $^{12}$CO(2--1) and $^{13}$CO(2--1) emission emission maps of
V4046 Sgr reveal a rotating disk around this binary
system extending out to a radius of $\sim$370~AU. A truncated
power-law fit to the $^{12}$CO(2--1) visibilities indicate the disk is
viewed at an inclination of 33$^\circ$ if the total central binary mass is
1.8~M$_\odot$ \citep[as determined by][]{sg04}. However, we cannot rule out the possibility that the
central binary mass is as small as 1.55~M$_\odot$ \citep[the value
determined by][]{quast00} on the basis
of the SMA CO data alone, since disk inclinations as large as
$\sim$40$^\circ$ (corresponding to a central mass of $\sim$1.4~M$_\odot$)
are not precluded by the modeling.

While the position angle of the central binary is unknown, the close
correspondence between the range of disk inclination determined from
the modeling (\S~4.2) and the inclination of the binary star orbit adopted
in previous binary system radial velocity studies
\citep[35$^\circ$;][]{quast00} leads us to conclude that the binary
pair and disk are nearly coplanar. The disk and orbital planes of
circumbinary disk systems are expected to be very similar, because the
system either formed with such an alignment or subsequently evolved
into an aligned state \citep{monin06}. It is therefore
significant that the disk and binary inclinations indeed appear to be
aligned to within a few degrees, in this $\sim$12 Myr-old system.

We resolved the 1.3 mm continuum emission from V4046 Sgr with the SMA,
and find that the flux is consistent with 37~K dust located
$\sim$40~AU from the central stars \citep{k08b}. We estimate a mass of
$\sim$40~Earth masses of dust in the disk, about a factor of 3 smaller than the
gas mass inferred from the CO line intensities ($\sim$110 Earth masses,
assuming a CO:H$_2$ ratio of 10$^{-4}$). Hence, either the residual
gas mass in the disk orbiting this rather ``old'' binary classical T
Tauri system is much smaller than that expected for
a gas-to-dust ratio of 100, or the disk is severely depleted in
gas-phase CO. Indeed, for the temperature profile adopted here, $T<20$
K for radii $\stackrel{>}{\sim}$200 AU, suggesting that much of the
CO present in the disk may be in the form of icy grain mantles.

Although it appears most protoplanetary disks have dissipated after
several million years \citep{haisch01,uzpen09}, V4046 Sgr, at age
$\sim$12 Myr, still retains a substantial amount of mass in its
circumbinary disk. Indeed, V4046 Sgr is one of only four known
pre-main sequence stars within 100~pc that possess molecular
disks. While V4046 Sgr is a close binary system, the other three
(TW~Hya, HD~141569, and 49~Ceti) do not appear to harbor close
companions. Previous submillimeter studies have revealed gas-rich
disks extending out to about 200~AU in both TW~Hya and 49~Ceti
\citep{qi04,hughes08} and $\sim$250~AU in HD~141569 \citep{Dent05}.
Our SMA imaging demonstrates that the V4046 Sgr disk (radius 370~AU)
is even larger and, when compared to the disk orbiting TW Hya (a
single star of slightly later spectral type), also retains a larger
gas mass (if we apply the same standard assumptions of CO 
abundance relative to H$_2$ of 10$^{-4}$ and $^{12}$C/$^{13}$C of 
89). The twin stars in V4046 Sgr evidently are still actively
accreting from this extensive, gaseous disk and, likewise, any young
Jovian planets within this system may be undergoing their final stages
of gas accretion.

  Giant planet formation in disks can occur in one of two ways:
  core accretion, with timescales of a few $\times10^6$~years
  \citep{Inaba03,Lissauer09} or disk instability, with timescales as
  short as a few $\times10^3$~years \citep{Boss00}. One
  expects that long-lived disks such as those around TW Hya and V4046
  Sgr (ages $\sim$10~Myr) will provide more opportunities to form
  giant planets than shorter-lived disks, regardless of how they form,
  so long as sufficient reservoirs of gas remain present. Meanwhile,
  studies have shown that disks around close ($<$100~AU) binary
  systems tend to dissipate faster than those around single stars or
  more widely separated systems \citep{Bouwman06,Cieza09}. Typically,
  therefore, if giant planets are to form around close binary systems,
  they must must form more rapidly than would be necessary for giant
  planets forming around single stars.  Indeed, there is evidence that planets 
  in binary systems with separations $<$100~AU are biased towards
  higher masses, possibly as a natural consequence of rapid planet formation
  via gravitational instability \citep{duchene09}.  

  V4046 Sgr therefore presents a particularly interesting and
  important case. Given its tight ($\sim$9~$R_\odot$) separation, the
  outer portions of its circumbinary disk, where giant planets are
  expected to form, will have been left relatively undisturbed
  throughout the system's lifetime. Only within $\sim$1~AU are
  dynamical
  effects due to the central binary expected to significantly
  affect the disk (\citealt{Jensen97} estimate that an inner hole out of
  radius $\sim$0.2~AU is necessary to account for the lack of excess
  infrared emission from V4046 Sgr shortward of 10~$\mu$m).
  While
  about 20\% of extrasolar planets have been detected in binary
  systems \citep{raghavan06,eggenberger07}, no unambiguous detection
  of an extrasolar planet has been made for a binary with separation
  $<$20~AU \citep[though the eclipse timings of HW Virginis and CM
  Draconis suggest that planets may orbit these two
  binaries;][]{deeg08,lee09}. Hence, there is ample motivation to continue to
scrutinize the V4046 Sgr disk at the highest possible spatial
resolution, to see if planets are already orbiting this binary system.

\acknowledgements {\it Acknowledgements.} We thank Ben Zuckerman for
useful discussions and comments and Meredith Hughes for lending us her 
deprojected visibility code. We are grateful to Michiel
Hogerheijde for providing access to the 2D version of RATRAN used in this
work. We appreciate the suggestions and comments of the anonymous
referee. The Submillimeter Array is a joint project between the
Smithsonian Astrophysical Observatory and the Academia Sinica
Institute of Astronomy and Astrophysics and is funded by the
Smithsonian Institution and the Academia Sinica. This research was
supported by NASA Astrophysics Data Analysis Program grant NNX09AC96G
to RIT and UCLA.

\clearpage

\begin{table}[h]
\caption{Observational Parameters}
\begin{center} {\footnotesize
\begin{tabular}{lcccc}
\hline
\hline
Parameter & $^{12}$CO(2--1) & $^{13}$CO(2--1) & C$^{18}$O(2--1) & Continuum \\
 & & & & 1.3mm \\
\hline
Rest Frequency (GHz) & 230.538 & 220.399 & 219.560 & 225\\
Beam Size (FWHM) & $2.24\arcsec \times 1.48\arcsec$ & $2.34\arcsec \times 1.66\arcsec$ & $2.35\arcsec \times 1.67\arcsec$ & $1.67\arcsec \times 1.01\arcsec$ \\
P.A. & $8^\circ$ & $4^\circ$ & $4^\circ$ & $1^\circ$ \\
RMS noise (mJy beam$^{-1}$) & 87 & 76 & 76 & 0.8\\
Peak Flux Density (mJy beam$^{-1}$) & 2640 & 790 & $<$300 & 210.5 \\ 
Integrated Line Intensity (Jy km s$^{-1}$) & 30.2 & 8.7 & $<$1 & -- \\ 
Integrated Continuum Flux (mJy) & -- & -- & -- & 360\\
\hline
\end{tabular} }
\end{center}
\label{tab1}
\end{table}

\begin{table}[htdp]
\caption{Model Properties}
\begin{center}
\begin{tabular}{ l l }
\hline
\hline
Parameter & Value \\
\hline
Central mass (M$_\odot$) & $1.8^a$ \\
Inclination (\arcdeg) & $33.4\pm0.3$ \\
Systemic LSR Velocity (km s$^{-1}$) & $2.92\pm0.01$ \\
P.A. (\arcdeg) & $76.4\pm0.5$ \\
$R_{in} (AU)$ & $4^{b}$\\
$R_{out}$ (AU) & $370\pm30$ \\
$n_{100}$ (cm$^{-3}$) & $4\times10^8$ $^b$ \\
$p$ & $2^b$ \\
$T_{100}$ (K) & $28^b$ \\
$q$ & $0.6^b$ \\
\hline
\end{tabular}
\end{center}
{\small $^a$See \S~\ref{massvary}\\
$^b$Adopted values; these are not fit. \\
Quoted uncertainties are 1$\sigma$. See text for more details.}
\label{tab2}
\end{table}

\end{document}